\pdfoutput=1

\documentclass[iop]{emulateapj}

\usepackage{natbib,epsfig,url,graphicx,amsmath,footnote,float,xcolor,cases}
\usepackage[colorlinks=true,linkcolor=blue,citecolor=blue]{hyperref}

\begin{document}

\submitted{AJ; accepted}

\title{Dynamical Constraints on Mercury's Collisional Origin}

\author{Matthew S. Clement\altaffilmark{1,*}, Nathan A. Kaib\altaffilmark{1}, \& John E. Chambers\altaffilmark{2}}

\altaffiltext{1}{HL Dodge Department of Physics Astronomy, University of Oklahoma, Norman, OK 73019, USA}
\altaffiltext{2}{Department of Terrestrial Magnetism, Carnegie Institution
for Science, 5241 Broad Branch Road, NW, Washington, DC
20015, USA}

\altaffiltext{*}{corresponding author email: matt.clement@ou.edu}

\setcounter{footnote}{0}
\begin{abstract}
	Of the solar system's four terrestrial planets, the origin of Mercury is perhaps the most mysterious.  Modern numerical simulations designed to model the dynamics of terrestrial planet formation systematically fail to replicate Mercury; which possesses just 5$\%$ the mass of Earth and the highest orbital eccentricity and inclination among the planets.  However, Mercury's  large iron-rich core and low volatile inventory stand out among the inner planets, and seem to imply a violent collisional origin.  Because most  algorithms used for simulating terrestrial accretion do not consider the effects of collisional fragmentation, it has been difficult to test these collisional hypotheses within the larger context of planet formation.  Here, we analyze a large suite of terrestrial accretion models that account for the fragmentation of colliding bodies.  We find that planets with core mass fractions boosted as a result of repeated hit-and-run collisions are produced in 90$\%$ of our simulations.  While many of these planets are similar to Mercury in mass, they rarely lie on Mercury-like orbits.  Furthermore, we perform an additional batch of simulations designed to specifically test the single giant impact origin scenario.  We find less than a 1$\%$ probability of simultaneously replicating the Mercury-Venus dynamical spacing and the terrestrial system's degree of orbital excitation after such an event.  While dynamical models have made great strides in understanding Mars' low mass, their inability to form accurate Mercury analogs remains a glaring problem.
\break
\break
{\bf Keywords:} Mercury, Planetary Formation, Terrestrial Planets, Collisional Fragmentation
\end{abstract}

\section{Introduction}

The late stages of terrestrial accretion (giant impact phase) have been studied numerically  \citep{duncan98,chambers99} by numerous authors.  By assuming a surface density of solids roughly commensurate with the minimum mass solar nebula \citep{mmsn}, numerical integrators produce appropriately massed Earth and Venus analogues (at the proper orbital locations) with great frequency regardless of whether other parameters are varied \citep{chambers01,obrien06,ray06,ray09a,hansen09,clement18}.  However, replicating the small mass of Mars ($\sim$10$\%$ that of the Earth) requires modification to the classic initial conditions \citep{wetherill91}; the exact mechanism for which is a topic of continued debate \citep{thommes08,walsh11,iz14,fischer14,draz16,ray17sci,clement18}.  Notably, \citet{ray09a} demonstrated that a configuration of Jupiter and Saturn  with a mutual inclination of 1.5 $^{\circ}$ and $e_{J}$ = $e_{S}$ = 0.1 (extra eccentric Jupiter and Saturn, EEJS) consistently produced a small Mars. However, these initial conditions appeared unlikely since interactions between the primordial gas disk and the growing gas planets are thought to result in a system of fully formed giant planets on near circular orbits \citep{papaloizou00,tanaka04,cresswell07}.  In a similar manner, the giant planets can influence Mars' formation via resonance sweeping (the so-called ``dynamical shakeup;'' \citet{thommes08,bromley17}) or mean motion resonance (MMR) crossing \citep{lykawaka13}.  It has also been proposed that the Mars forming region and primordial asteroid belt were largely empty, and subsequently implanted with material from the outer solar system during Jupiter's growth phase \citep{izidoro15,ray17sci,ray17}.  An alternative idea, the ``Grand Tack'' hypothesis \citep{walsh11,walsh16}, suggests that the still-forming Jupiter and Saturn migrated in to the inner solar system during the gas disk phase of evolution, and subsequently back out.  By truncating the primordial disk of planetesimals at $\sim$ 1 au, the terrestrial planets form out of a narrow annulus, and Mars' final mass is greatly limited \citep{hansen09}.  Finally, a giant planet orbital instability (the so-called ``Nice Model;'' \citet{gomes05,Tsi05,mor05}) timed in conjunction with the late stages of terrestrial planet formation can also result in a small Mars \citep{clement18,clement18_frag,clement18_ab}.  Interestingly, the Nice Model instability might also explain Mercury's uniquely excited orbit \citep{roig16}.

Despite these significant advances, fully understanding the formation of Mercury (which is only $\sim$5$\%$ the mass of Earth) remains a mystery to dynamicists.  Most previous studies of terrestrial planet formation find that Mercury analogues form  in only 5-10$\%$ of simulated systems \citep{chambers01,ray09a,clement18}.  While terrestrial planet formation simulations occasionally produce planets near Mercury's location, its other physical characteristics seem to suggest it also had a unique accretion history.  In particular, it has an unusually large iron-rich core ($\sim$70-80$\%$ of its entire mass; \citet{merc_core,hauck13,mercury_review,margot18}) and depleted volatile inventory (however the MESSENGER mission detected abundances of moderately volatile elements similar to that of the other terrestrial planets).  These features have generally been interpreted as the consequences of an energetic, fragmenting collision during its formation \citep{benz88,asphaug10,asphaug14}, though other explanations have also been proposed.  

Models of terrestrial planet formation in the solar system require truncation of the inner disk at $\sim$0.5 au in order to replicate the modern mass distribution and orbital excitation profile of the inner planets.  However, the number of known multi-planet systems of terrestrial exoplanets with semi-major axes less than Mercury's (eg: Kepler-102, Kepler-11, Kepler-85 and TRAPPIST-1; \citet{lissauer13,lissauer14,gillon17}) seem to imply that this is not always the case.  In fact, \citet{mulders18} argued that as many as $\sim$90$\%$ of Main Sequence stars possess close-in Super-Earths.  \citet{volk15} proposed that the young solar system may have contained a similar population of tightly packed planets.  In this scenario, some time in the solar system's early history these planets underwent a cataclysmic instability that left behind Mercury as the lone survivor.  However, the simulations of \citet{volk15} demonstrated quasi-stability in known systems of exoplanets and did not include realistic collisions (see further discussion in \citet{wallace17}).  Furthermore, the systems studied in that work were not necessarily analogous to the young solar system since they did not include the other planets.  In an alternative scheme, the lack of planets interior to Mercury might be explained by planetesimal shepherding during Jupiter's outward migration phase \citep{ray16}.  Another idea is that Jupiter's inward migration in the Grand Tack model could have  cleared out the Mercury forming region via resonant excitation of small planetesimals \citep{batygin15}, though this scenario directly conflicts with the results of accretion models \citep{kenyon09,ray16}.    

Nevertheless, Mercury's iron-rich composition makes the collisional hypothesis a compelling one.  However, most modern integration schemes \citep{duncan98,chambers99,genga} used for modeling the giant impact phase of terrestrial accretion treat all collisions as perfectly accretionary.  Thus, if Mercury truly originated from a large, energetic collision during the tail-end of planet formation, it makes sense that N-body simulations consistently fail to produce accurate Mercury analogs.  For recent reviews on the topics of Mercury's formation and composition consult \citet{ebel17}, \citet{nittler17}and \citet{margot18}.

While the literature concerning the collisional origin of the moon is extensive \citep{hartmann75,cuk12,canup12,herwartz14,mast15,kaibcowan15,quarles15,rufu17,citron18}, the topic of Mercury's collisional origin is not as well explored.  Nevertheless, two different collisional scenarios \citep{ls14,ebel17} have been proposed to explain Mercury's high core mass fraction (CMF) and low concentration of volatiles.  The first scheme involves the erosion of an originally larger planet by repeated hit-and-run collisions \citep{asphaug06,svetsov11,chau18}.  However, hydrodynamical simulations indicate that the disrupted volatiles would eventually be re-accreted, and the resulting planet would not be as volatile poor as Mercury \citep{marchi14}.  The second scenario concerns a fragmenting collision between two large protoplanets \citep{benz07,asphaug10,asphaug14,chau18}.  While the logical target for a Mercury-forming impact would be the proto-Venus, the lack of an internally generated magnetic dynamo might be evidence that primordial stratification still exists within the Venus' core.  This may imply that the planet was never disrupted by such a large impact \citep{jacobson17b}.

While the specific collisional parameters required in either Mercury-forming scenario have been evaluated before \citep{asphaug06,svetsov11,asphaug14,chau18}, the full picture of the event within the greater context of terrestrial planet formation as a whole is not as well understood.  Since the various regimes of collisional parameter space have been mapped by \citet{lands12} and \citet{genda12}, traditional N-body integrators used for investigating the dynamics of terrestrial planet formation can be  modified to include the effects of collisional fragmentation.  \citet{chambers13} performed the first such simulations, and found several examples of the fragmentation processes producing planets with CMFs similar to Mercury.  \citet{dwyer15} performed an in-depth analysis of the bulk chemical content of the planets formed in these same simulations, and noted that it was difficult to form Mercury-like planets with low silicate mass fractions.  However, the small sample size of just 8 integrations makes it difficult to draw any statistical conclusions from these simulations.  Additional authors have used similar N-body codes to study planetary dynamics \citep{bonsor15,carter15,leinhardt15,quintana15,wallace17}.  However, no dedicated study of the dynamical barriers involved in forming Mercury via repeated hit-and-run collisions or a single giant impact has been performed.  It should also be noted that \citet{lykawka17} investigated the possibility of Mercury forming naturally in the inner region (between 0.2 and 0.5 au) of the terrestrial disk, and found that the final Mercury analogs were typically over-massed and under-excited.  In this paper, we systematically investigate Mercury's collisional origin using the modified version of the $Mercury6$ hybrid integrator \citep{chambers99} described in \citet{chambers13}.  Our analysis is divided into three segments.  First, we investigate the frequency of planets with Mercury-like compositions formed in a large sample of 360 full simulations of terrestrial planet formation.  Next, we analyze the architectures of these systems of terrestrial planets, and specifically look for Mercury-like planets on Mercury-like orbits.  In the final section of this manuscript, we comment on the dynamical barriers involved in the scenario where Mercury forms via a single giant impact.

\section{Methods}

We investigate three suites of simulations designed to replicate the early instability scenario  \citep{clement18,clement18_frag,clement18_ab}, the Grand Tack and low mass asteroid belt hypotheses \citep{walsh11,ray17sci}, and a control set using classic initial conditions \citep{chambers01,chambers13}.  For a complete discussion and full presentation of these simulations, consult \citet{clement18_frag}. 

Our work utilizes the same modified version of the $Mercury6$ hybrid integrator \citep{chambers99} described in \citet{chambers13} and \citet{dwyer15}.  The collisional fragmentation algorithm functions by first calculating the largest remnant utilizing relations from \citet{lands12}.  The leftover mass is divided in to an appropriate number of equal-massed fragments, with masses greater than the minimum fragment mass (MFM).   Our simulations use a 6 day time-step and set the MFM to $\sim$10$\%$ the mass of Mercury.  Because collisions producing $\gtrsim$ 90 fragments can cause the Bulirsch-Stoer portion of $\it{Mercury's}$ hybrid-symplectic integrator to be bogged down, we find a MFM of 0.0055 $M_{\oplus}$ to be the limit of our resolution (for a more in depth discussion of this see \citet{wallace17}).  Objects are considered to be merged with the Sun at 0.1 au, and removed via ejection at 100 au.  Because of the hybrid symplectic scheme's inability to accurately integrate through low pericenter passages \citep{chambers99}, removing objects that pass within $\sim$0.1 au of the Sun is common practice in similar numerical studies \citep{chambers01,chambers13}.

\subsection{Control Simulations}
For our primordial control disks, we choose the simplest set of initial conditions consistent with those chosen by many previous authors \citep{chambers01,ray06,obrien06,chambers13,kaibcham16,clement18}.  We first begin 100 integrations of a 5 $M_{\oplus}$ terrestrial forming disk of bodies distributed between 0.5 and 4.0 au.  As in \citet{clement18}, half the disk mass is placed in 100 equal-mass embryos, and the other half in 1,000 equal-mass planetesimals.  While the planetesimals do not interact gravitationally or collide with one another, they can experience fragmenting collisions if they collide with one of the embryos.  When this is the case, the resulting fragments are also treated as planetesimals (fragments from embryo-embryo collisions are considered fully self-gravitating).  The radial spacing within the disk is selected to achieve a surface density profile proportional to  $r^{-3/2}$ \citep{birnstiel12}.  Angular orbital elements are drawn randomly, and eccentricities and inclinations are selected from near circular gaussian distributions  ($\sigma_{e}=.02$ and $\sigma_{i}=.2^{\circ}$).  Thus all initial eccentricities and inclinations are below 0.001 and $1^{\circ}$, respectively.  Since the outer planets are thought to form first \citep{haisch01,halliday08,kleine09}, we include Jupiter and Saturn in a 3:2 MMR ($a_{J}$ = 5.6 au, $a_{S}$ = 7.6 au) in these integrations \citep{wetherill96,chambers_cassen02,levison_agnor03,raymond04,clement18}.  These simulations are evolved for 200 Myr, and become our control set.

\subsection{Instability Scenario}
To test the effects of a Nice Model style instability \citep{gomes05,Tsi05,mor05}, we take snapshots of our control disks at 1,5 and 10 Myr, and input them in to unstable giant planet configurations \citep{nesvorny11,clement18}.  Because ice giants are routinely excited to the point of ejection in simulations of the classical Nice Model \citep{gomes05}, the current version includes one or two additional primordial ice giants \citep{nesvorny12}.  As in \citet{clement18}, we perform one set of integrations using a 5 giant planet configuration, and a second with a 6 giant planet configuration.  These simulations are evolved for an additional 200 Myr using the same integrator and time-step described above.  To ensure we only sample systems where the evolution of the giant planets is most akin to the actual solar system, simulations that fail to eject an ice giant within 5 Myr, and those where Jupiter and Saturn's period ratio exceeds 2.8 (the present ratio is 2.49) are deleted.  Through this process, our initial sample of 600 instability simulations is reduced to 160.  

There are advantages and disadvantages to both a 5 and 6 giant planet instability.  For instance, a 6 planet setup excites Jupiter's eccentricity in two stages (with each primordial ice giant ejection).  This evolutionary scheme can potentially prevent Jupiter's eccentricity from being damped to below its modern value via secular friction \citep{nesvorny12}.  Since the precise nature of the solar system's instability is unconstrained, we study both types of giant planet configurations (for a complete discussion of this methodology consult \citet{clement18} and \citet{clement18_frag}).  Other previous works test the effects of the Nice Model by performing a large number of instability simulations, selecting the run with the best final giant planet architecture, and then ``replaying'' the chosen instability in the presence of the objects of interest to the study (eg: the terrestrial planets, asteroid belt, etc; \citet{brasil16,roig16,deienno17}).  However, given the unconstrained, chaotic nature of the giant planets' particular evolution within the instability, we choose to perform many simulations, and then select a sample of the best final Jupiter-Saturn configurations \citep{kaibcham16,clement18}.

\subsection{Annulus Scenario}
To roughly replicate the initial conditions supposed by the Grand Tack \citep{walsh11} and low mass asteroid belt \citep{ray17sci} models, we lay down 400 equal-mass planet embryos in a narrow annulus between 0.7 and 1.0 au \citep{hansen09}.  Planetesimals are not used in these 100 simulations, and thus each object interacts gravitationally with every other object in the run), and the total disk mass is set to 2.0 $M_{\oplus}$.  The integrator, time-step and method of selecting orbital elements are the same as described above.  Including the additional 100 control simulations and 160 instability runs, our total sample of fully accreted systems of terrestrial planets is 360.  A summary of our three sets of terrestrial planet formation simulations is provided in table \ref{table:ics2}.

\begin{table*}
\centering
\begin{tabular}{c c c c c c c}
\hline
Set & $a_{in}$ (au) & $a_{out}$ (au) & $M_{tot}$ ($M_{\oplus}$) & $N_{emb}$ & $N_{pln}$\\
\hline
Control & 0.5 & 4.0 & 5.0 & 100 & 1000 \\	
Instability & 0.5 & 4.0 & 5.0 & 100 & 1000 \\	
Annulus & 0.7 & 1.0 & 2.0 & 400 & 0 \\	
\hline	
\end{tabular}
\caption{Summary of initial conditions for complete sets of terrestrial planet formation simulations.  The columns are (1) the name of the simulation set, (2) the inner edge of the terrestrial forming disk, (3) the disk's outer edge, (4) the total disk mass, (5) the number of equal-mass embryos used, and (6) the number of equal-mass planetesimals used.}
\label{table:ics2}
\end{table*}

\subsection{Analysis Metrics}

\subsubsection{Core Mass Fraction Calculation}
In order to filter out Mercury-like objects with a high CMFs \citep{merc_core,hauck13}, we calculate the mass of each object's iron core using the same method described by \citet{chambers13}.  We first assume that each body is fully differentiated, with 30$\%$ of its mass concentrated in the core, and the other 70$\%$ representing its silicate-rich mantle material.  In fully accretionary collisions, the new object's core mass is equal to the sum of the impactor and target bodies' core masses.  In a fragmenting collision, the fragments come from the mantle material.  If the mantle is depleted, the remaining fragments come from the core material.  If the fragmenting collision is of the hit-and-run type, the fragments first come from the mantle of the projectile, and then from its core.  In the subsequent text, we refer to an object with CMF $>$ 0.5 as a ``high CMF'' object.

\subsubsection{Terrestrial Angular Momentum Deficit}

To compare the orbital excitation of our systems to the solar system quantitatively, we calculate the normalized angular momentum deficit (AMD, equation \ref{eqn:amd}) for all of our integrated systems \citep{laskar97}.  AMD quantifies the deviation of the orbits in a system from perfectly co-planar, circular orbits.  While the terrestrial planets' AMD can evolve marginally over Gyr timescales \citep{laskar97,agnor17}, a simulated system finishing with an AMD greater than twice the modern value of $AMD_{MVEM}$ is considered a poor solar system analog by most authors \citep{ray09a,clement18}.

\begin{equation}
	AMD = \frac{\sum_{i}m_{i}\sqrt{a_{i}}[1 - \sqrt{(1 - e_{i}^2)}\cos{i_{i}}]} {\sum_{i}m_{i}\sqrt{a_{i}}} 
	\label{eqn:amd}
\end{equation}

\subsubsection{Mercury-Venus dynamical spacing}

In general, a successful Mercury analog must undergo a series of CMF-reducing collisions during the planet-formation epoch, and then finish on an orbit that is stable for $\sim$4 Gyr.  Because late ($t>200$ Myr) instabilities are common results of terrestrial planet formation models \citep{clement17}, we are particularly interested in Mercury-like planets sufficiently dynamically separated from Venus-like planets (eg: not on crossing orbits).  For this reason, we compare the difference in the perihelion of Venus analogs and the aphelion of Mercury analogs, as well as the planets' relative period ratios, to solar system values throughout our study.  However, forming Mercury on an orbit well separated from Venus does not necessarily imply that the same is true for Venus and Earth, or Earth and Mars.  For that reason we also employ metrics from \citet{clement18} and \citet{clement18_frag} to quantify the entire inner solar system's structure.  To satisfy this criterion (criterion A), a terrestrial system must contain 4 planets meeting the following requirements: $M_{Merc}<$ 0.2 $M_{\oplus}$; $M_{Ven,Ear}>$ 0.6 $M_{\oplus}$; $M_{Mars}<$ 0.3 $M_{\oplus}$; $a_{Merc}<$ 0.5 au; 0.5 au $<a_{Ven,Ear}<$ 1.3 au; and 1.3 au $<a_{Mars}<$ 2.0 au.  Given the stochastic nature of the planet formation process, it is unreasonable to expect our systems simultaneously match a large number of highly specific constraints \citep{nesvorny12,clement18}.  Therefore, we maintain broad success criteria, and avoid multiplying metrics.

\section{Results and Discussion}

\subsection{High CMF Objects}

\begin{figure}
	\centering
	\includegraphics[width=.51\textwidth]{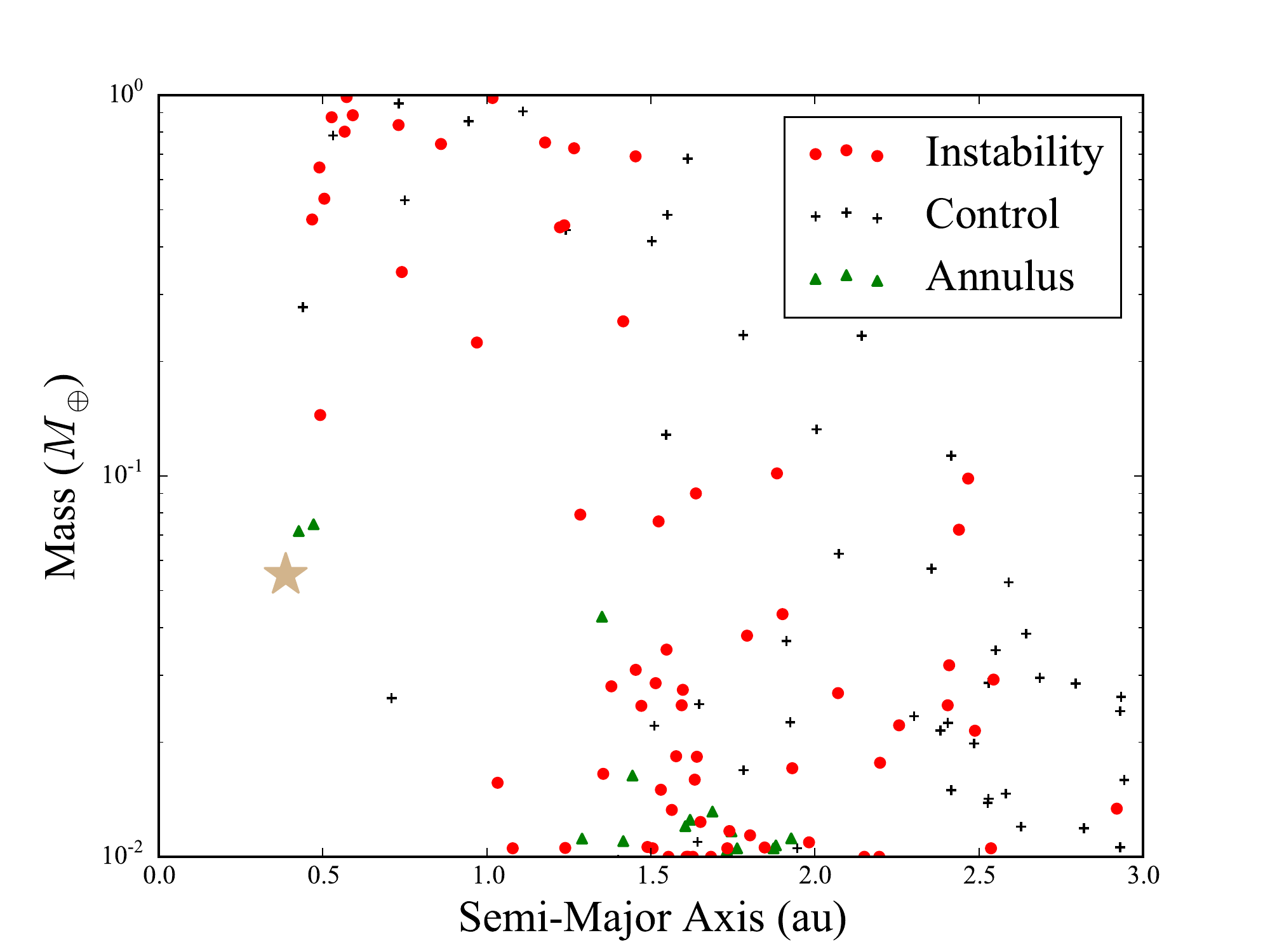}
	\caption{Distribution of all objects with masses greater than 0.01 $M_{\oplus}$ and CMFs greater than 0.5 surviving 200 Myr of terrestrial planet formation.  The colors black, red and green correspond to the control, instability and annulus simulations respectively.  The yellow star plots the actual planet Mercury.}
	\label{fig:cmf}
\end{figure}

The integrator's fragmentation routine ejects equal massed fragments in uniform directions along the collisional plane.  When an impact with sufficient energy to erode core material occurs, it is realistic to assume that the resulting fragments will not be made of equally mixed fractions of core and mantle material.  More likely, many fragments will be made entirely of mantle elements, and a few will come mostly from the core \citep{chambers13}.  Our method of assigning mantle material to the first fragments produced is plausible for embryos that have their CMFs boosted as the result of repeated hit-and-run collisions.  However, when a particularly energetic collision creates many fragments, and fully erodes the entire mantle inventory, we must randomly assign particular fragments as originating from the core.  In figure \ref{fig:cmf}, we plot all objects larger than 0.01 $M_{\oplus}$ from all of our different batches of terrestrial planet formation simulations with CMFs greater than 0.5.  While the annulus simulations produce the best Mercury analogs, the instability simulations prove most efficient at producing very large objects with high CMFs (several of which are nearly the mass of the Earth).  Since the average impact velocities are higher in the annulus set  due to the higher initial surface density and lower embryo masses ($v_{imp}/v_{esc}=$ 2.51 for fragmenting collisions as compared to 2.14 and 2.26 for the control and instability sets, respectively), the fragmentation process is more likely to grind high CMF objects down in to many, smaller particles.  This is also evidenced by the larger percentage of hit and run collisions (42$\%$ of all collisions) in the annulus set versus the other sets (22$\%$ for control and 29$\%$ for instability).  Indeed, our fully formed annulus simulations contain approximately twice as many objects with masses less than Mercury and perihelia less than 2.0 au than our other simulation sets.  Thus, our results imply that the higher initial surface density of the annulus set makes it difficult for those systems to produce high massed, high CMF objects via the multiple hit-and-run process.  Similarly, the instability batch is less efficient at producing lower massed, high CMF objects. 

The majority of the larger objects in figure \ref{fig:cmf} are embryos with CMFs boosted as a result of repeated hit-and-run collisions.  Smaller collisional fragments with masses closer to the MFM are more common, but also more likely to be on unstable orbits.  The reason for this is that our sample is biased towards objects that are produced near the end of our integration.  Many high CMF objects are produced throughout the duration of the simulation that eventually merge with one of the growing terrestrial planets, or are lost from the system via ejection or merger with the Sun.  As the terrestrial planets grow larger over the course of the simulation, the MFM is smaller relative to the the total mass involved in fragmenting collisions, and more fragments are produced.  Because of this, and the fact that we randomly assign mantle and core material to fragments, we are also interested in the high CMF objects that were produced, but eventually lost (but may have survived if assigned to a different fragment).  We note that the early instability scenario \citep{clement18,clement18_frag,clement18_ab} produces the most objects with boosted CMFs per simulation.  This is the result of the orbital excitation induced in the inner solar system by the chaotic perturbations from the unstable giant planets.  Indeed, fragmenting collisions account for $\sim$11$\%$ of all collisions in the instability runs.

Nevertheless, high CMF objects are a common type of object produced in our planet formation simulations, regardless of the initial conditions.  We note many examples of such bodies surviving the planet formation process on stable orbits.  90$\%$ of all our simulations finish with at least one high CMF object.  However, some of these bodies that finish in the asteroid belt and Mars region are more representative of small debris, leftover from the terrestrial formation process that will be eventually cleared in the subsequent 4.5 Gyr of evolution (see \citet{chambers13} and \citet{clement18_frag} for a further discussion of the longer accretion timescales in fragmentation simulations).

\subsection{Prevalence of Mercury Analogs}

The CMF results presented in the previous section should be taken in context with the fact that Mercury analogs (regardless of CMF) are still rare in all of our simulations.   We define a Mercury analog as any planet with m $<$ 0.2 $M_{\oplus}$ and $a<$ 0.5 au.  Our most successful simulation set at meeting this metric is the annulus set (16$\%$).  However, only two of those Mercury analogs have CMFs greater than 0.5.  Both the instability scenario and control runs meet this metric less than 2$\%$ of the time.  Furthermore, only one of our criterion A \citep{clement18} satisfying simulations yields a high CMF Mercury analog.

\begin{figure}
	\centering
	\includegraphics[width=.51\textwidth]{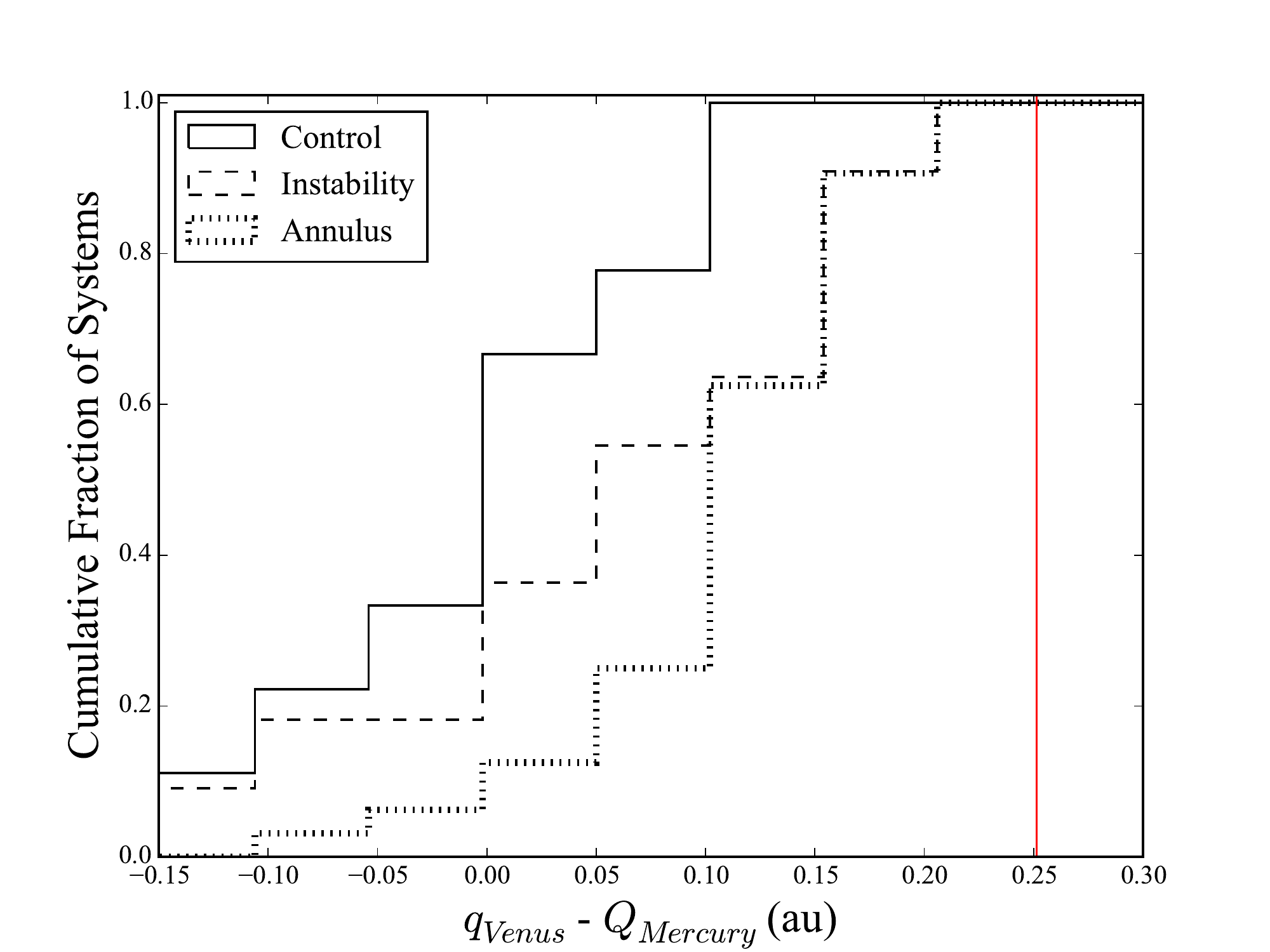}
	\qquad
	\includegraphics[width=.51\textwidth]{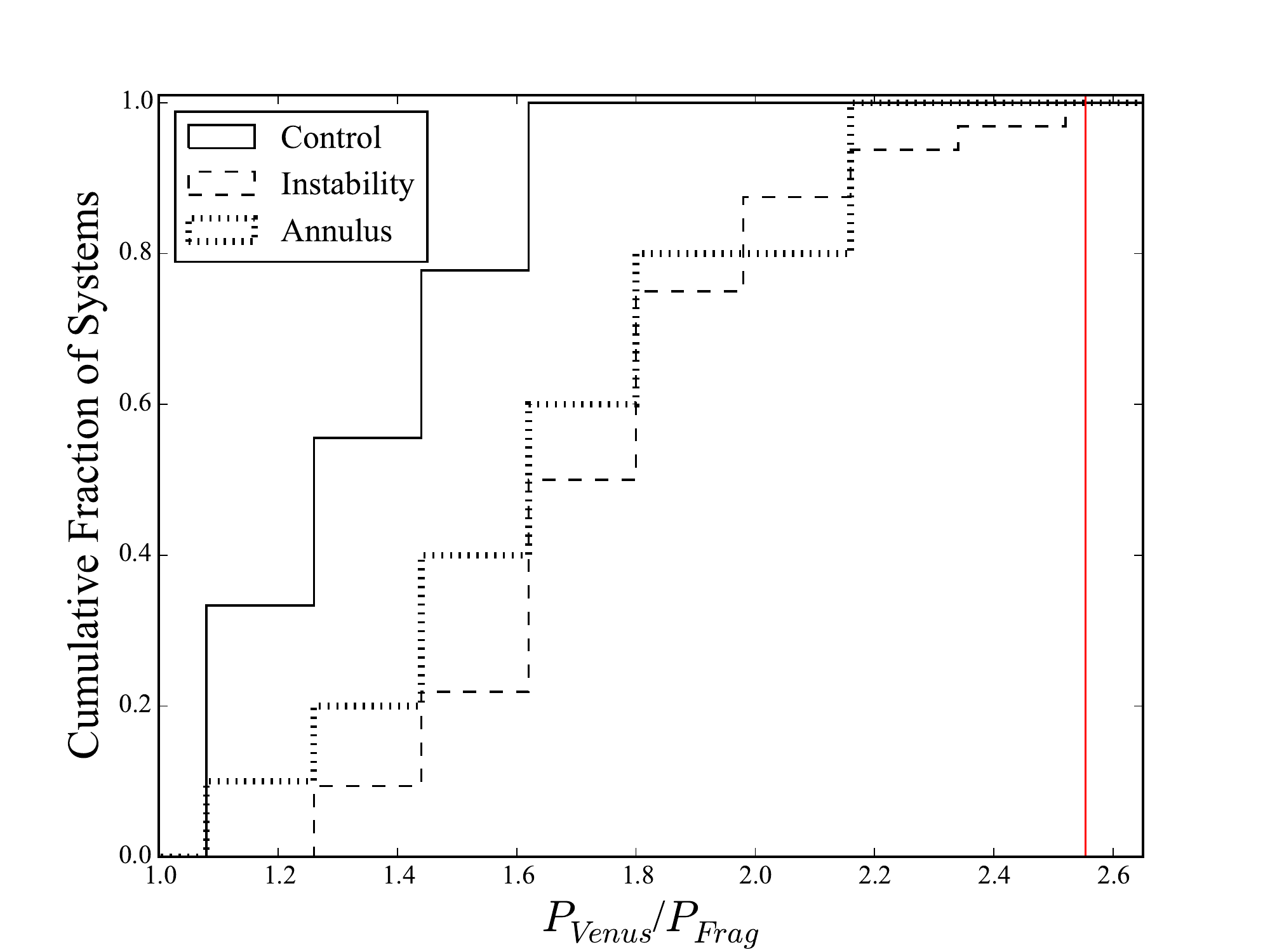}
	\caption{Cumulative distribution of the difference between the perihelion of Venus analogs and aphelion of Mercury analogs (top panel), and relative period ratios (bottom panel), produced in our complete simulations of terrestrial planet formation.  The figure only depicts systems that finish with a planet (Mercury) with $a<$ 0.5 au and m $<$ 0.2 $M_{\oplus}$, interior to a larger body (Venus) with $a<$ 1.3 au and m $>$ 0.6 $M_{\oplus}$.  The red vertical lines corresponds to the solar system values for Mercury and Venus.  The different line styles denote our different simulation sets.}
	\label{fig:qdelt_form}
\end{figure}

  In the actual solar system the difference between Mercury's aphelion and Venus' perihelion is $\sim$0.25 au.  Figure \ref{fig:qdelt_form} plots the cumulative distribution of $q_{Venus}-Q_{Mercury}$ for all systems that form a Mercury-Venus pair (planet with a $<$ 0.5 au and m $<$ 0.2 $M_{\oplus}$, interior to a larger body with a $<$ 1.3 au and m $>$ 0.6 $M_{\oplus}$).  It is clear that systems with Mercury-Venus spacings similar to the solar system are almost non-existent in all 3 of our simulation sets.  This is largely due to our chosen initial conditions (specifically the location of the inner terrestrial disk's edge).  Truncating the primordial terrestrial disk at 0.5 or 0.7 au is often justified in the literature as a means of preventing Earth or Super-Earth massed planets from growing near Mercury's orbit \citep{chambers01,ray09a}.  These authors also cite Mercury's hypothetical collisional origin \citep{benz88} as a rationale for neglecting the planet in N-body studies of planet formation in the solar system \citep{ray09a}.  Our work indicates that such assumptions are precarious from a dynamical standpoint.  Even in our most successful simulation set (annulus), where 16$\%$ of systems form small (m $<$ 0.2 $M_{\oplus}$) planets interior to 0.5 au, only 2 such systems meet criterion A, and no system finishes with $P_{Venus}/P_{Mercury}$ within 20$\%$ of the solar system value.

\begin{figure*}
	\centering
	\includegraphics[width=.8\textwidth]{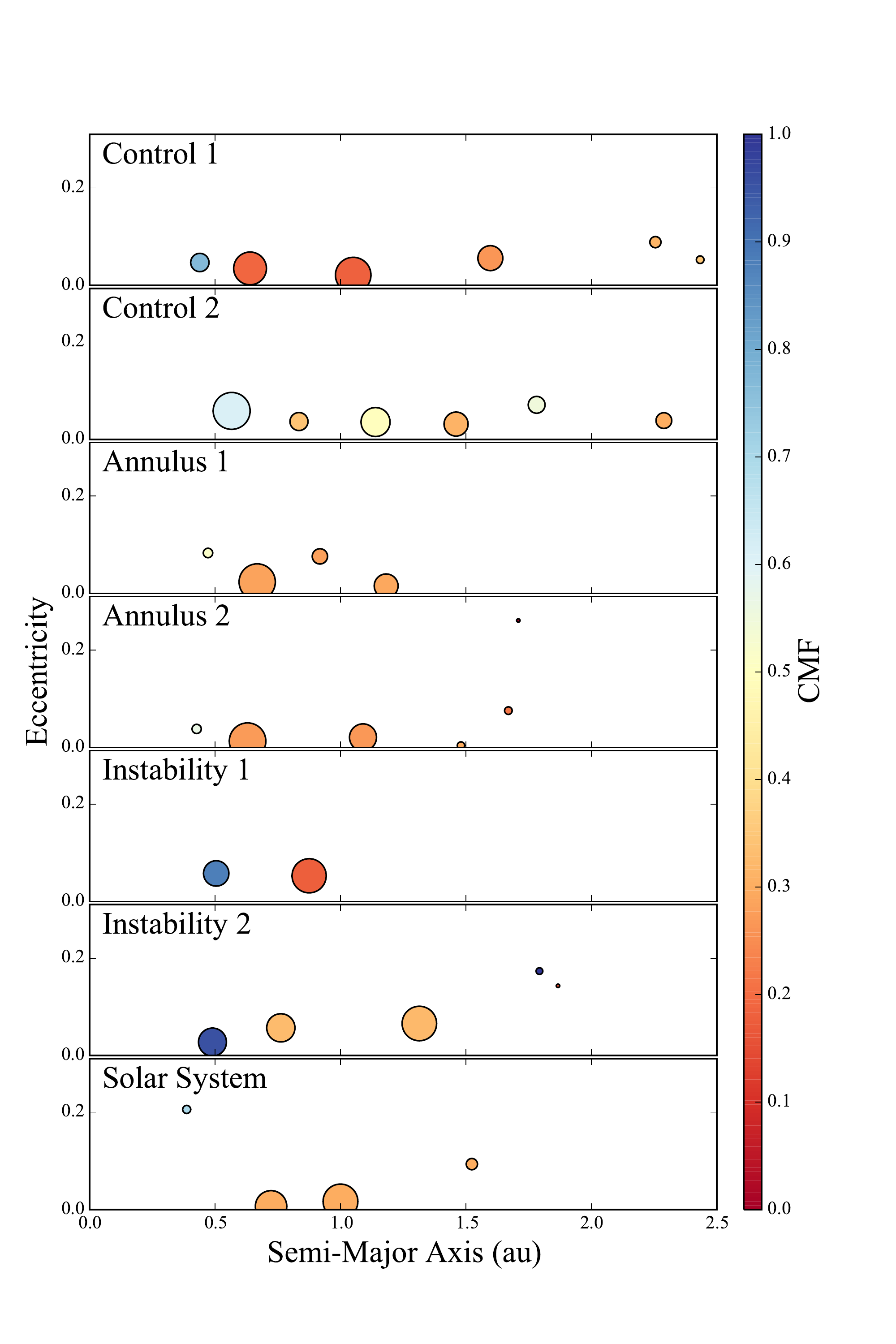}
	\caption{Semi-Major Axis/Eccentricity plot of selected successful simulations.  The size of each point is proportional to the mass of the planet.  The color of each point indicates the planet's CMF.  The top 6 panels show examples from each of our 3 simulation sets (Control, Annulus and Instability).  The bottom panel shows the actual solar system.}
	\label{fig:combined}
\end{figure*}

\begin{figure}
	\centering
	\includegraphics[width=.51\textwidth]{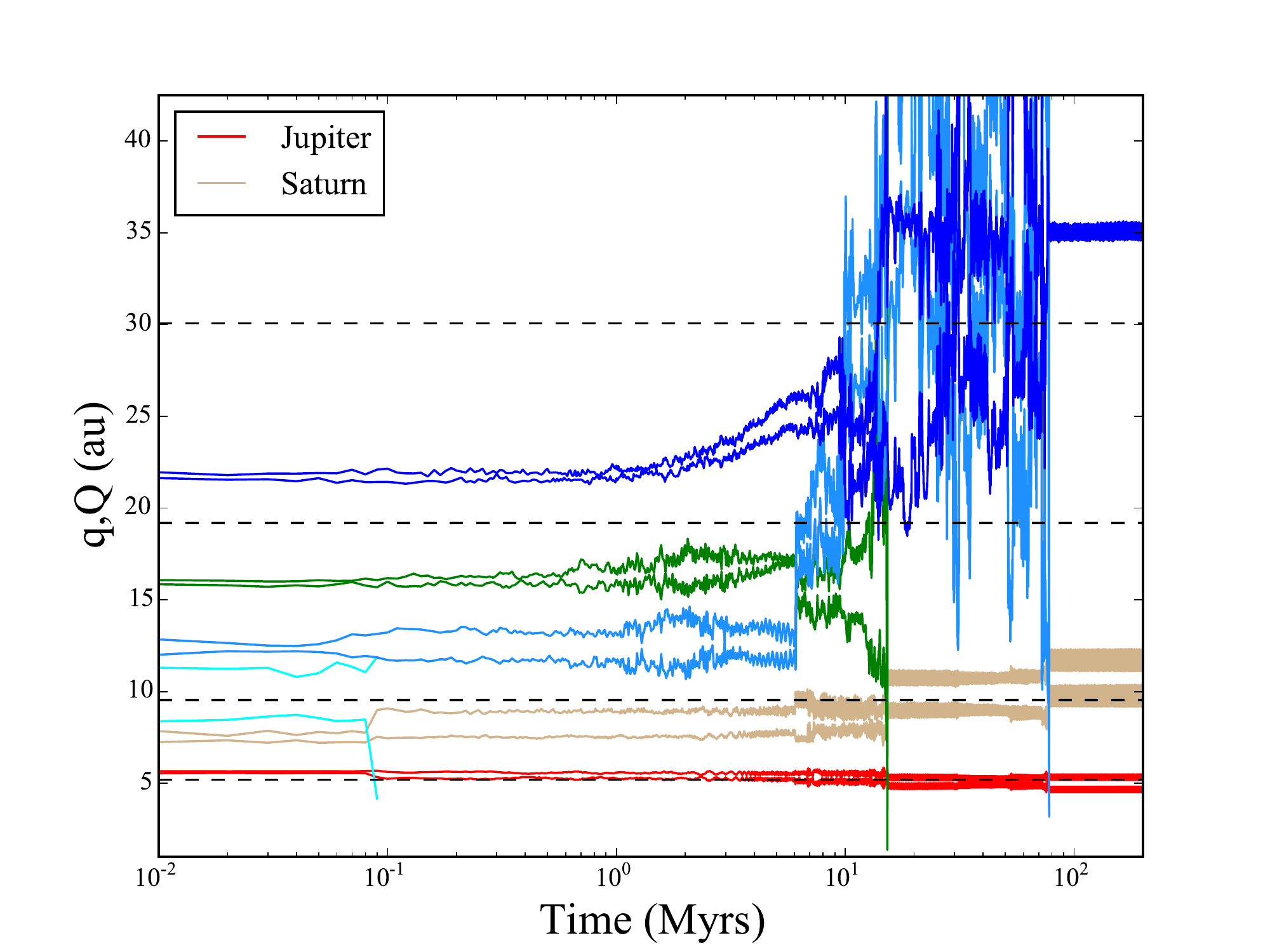}
	\includegraphics[width=.51\textwidth]{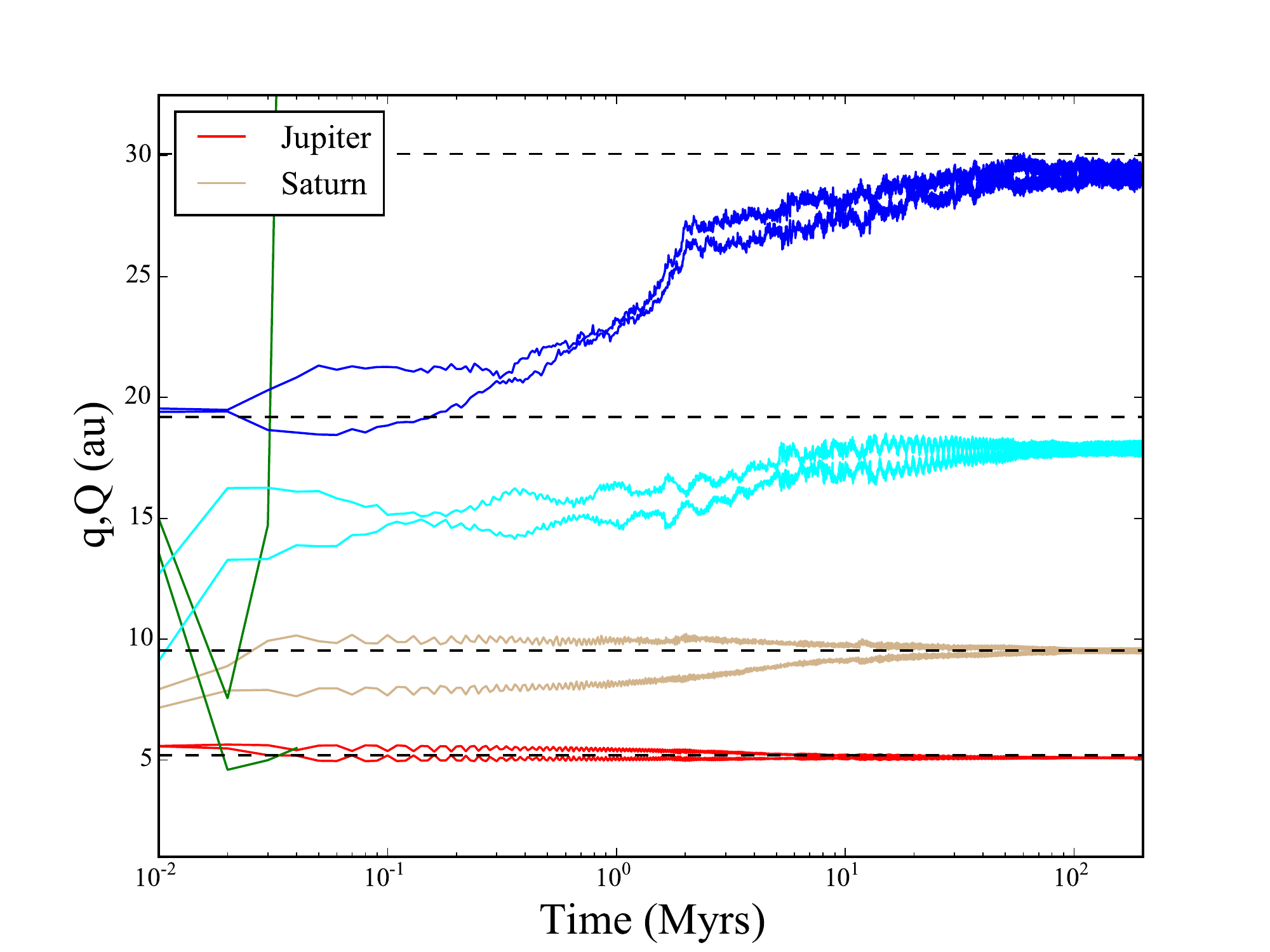}
	\caption{Giant planet instability evolutionary schemes for the two instability systems from figure \ref{fig:combined}.  The perihelia and aphelia are plotted for each outer planet in the simulation.  The horizontal dashed lines correspond to the modern semi-major axes of Jupiter, Saturn, Uranus and Neptune.  The top panel corresponds to ``Instability 1'' and the bottom panel plots ``Instability 2.''}
	\label{fig:Qaq2}
\end{figure}

In figure \ref{fig:combined}, we plot examples of final inner solar system architectures where a high CMF object finishes on a Mercury-like orbit.  While none of these systems match all aspects of the actual solar system, they demonstrate that the fragmentation process within the larger context of terrestrial planet formation is a viable explanation for Mercury's peculiar composition.  In particular, the system denoted ``Annulus 2'' provides an excellent match to the real masses and CMFs of Mercury, Venus, Earth and Mars.  As is true in the vast majority of our simulations, however, the semi-major axis spacing of the inner planets in this system is incorrect.  We also provide plots of the giant planets' evolution within the Nice Model for figure \ref{fig:combined}'s two Instability simulations in figure \ref{fig:Qaq2}.  Though neither final giant planet configuration perfectly matches the solar system, both Jupiter-Saturn systems finish in a state mostly analogous to the real one.  Because the terrestrial disk is less affected by the particular dynamics of the ice giants \citep{clement18}, we consider both of these outcomes adequate for our analysis.

\subsection{Dynamical Barriers to the Single Impact Scenario}

Our terrestrial accretion simulations rely on the stochasticity of the planet formation process to generate high CMF planets similar to Mercury.  Such planets have their mantle inventories depleted via complex and unique sequences of fragmenting collisions, erosive hit-and-run impacts, and accretion events with objects also altered in CMF.  Taking in to account the chaotic nature of planet formation, and the fact that $\sim$90$\%$ of our simulations finish with a high CMF terrestrial planet, it seems reasonable to argue that this process might explain Mercury.  However, similar to the results of previous studies \citep{chambers01,ray09a,clement18}, our simulations consistently fail to generate Mercury-massed planets sufficiently dynamically separated from Venus-like planets.  In this final section, we perform an additional suite of simplified simulations to study whether a single giant impact, occurring at the end of the planet formation epoch, might explain Mercury's offset from Venus.

Our simulations focus on the preferred single energetic impact scenario from \citet{asphaug14}, where a 0.25 $M_{\oplus}$ object strikes a 0.85 $M_{\oplus}$ target.  \citet{chau18} also agreed with this most promising scheme in a similar study that considered multiple initial target and projectile masses and several different collisional scenarios.  At the beginning of such a scenario, we assume that the potential targets are nearly formed versions of Venus, Earth ($M_{Earth}=M_{Venus}=$ 0.85 $M_{\oplus}$), and Mars on their modern orbits, while the projectile occupies an unstable orbit. To determine suitable unstable locations to place the projectile, we first use the WHFAST integrator \citep{whfast} in the REBOUND simulation package \citep{rebound} to investigate the object's stability in a/e space.  To account for the effects of general relativity, we utilize the additional forces provided in the expanded REBOUNDx library \citep{reboundx}.  We probe all regions of $a/e$ phase space (0.1$<a<$2.0 au) with 50 Myr integrations.  It should be noted that 50 Myr may not be long enough to detect unstable regions because secular resonance overlaps can take Gyr timescale to develop \citep{lithwick11}.  In the interest of minimizing computational time, we chose 50 Myr for a first-order approximation of the parameter space.  

After mapping this parameter space, we designate 4 different zones in which to place a potential Mercury-forming projectile. These are listed in table \ref{table:ics}.  We then perform a large suite of 1 Gyr simulations of these scenarios using a 5 day time-step.  Each integration includes the effects of general relativity, and sets the MFM to 0.025 $M_{\oplus}$.  In the first 3 scenarios, we vary the speed at which the fragments are ejected in random directions within the collisional plane (5, 10 and 20$\%$ greater than the mutual escape velocity).  In Scenario 4, we evaluate the high inclination parameter space proposed in \citet{jackson18}.

Regardless of the parameters varied, accurate Mercury analogs are exceedingly rare in this suite of single giant impact simulations.  In the following subsections, we briefly discuss the scenario's major shortcomings.

\subsubsection{Dynamical Offset from Venus}

\begin{figure}
	\centering
	\includegraphics[width=.51\textwidth]{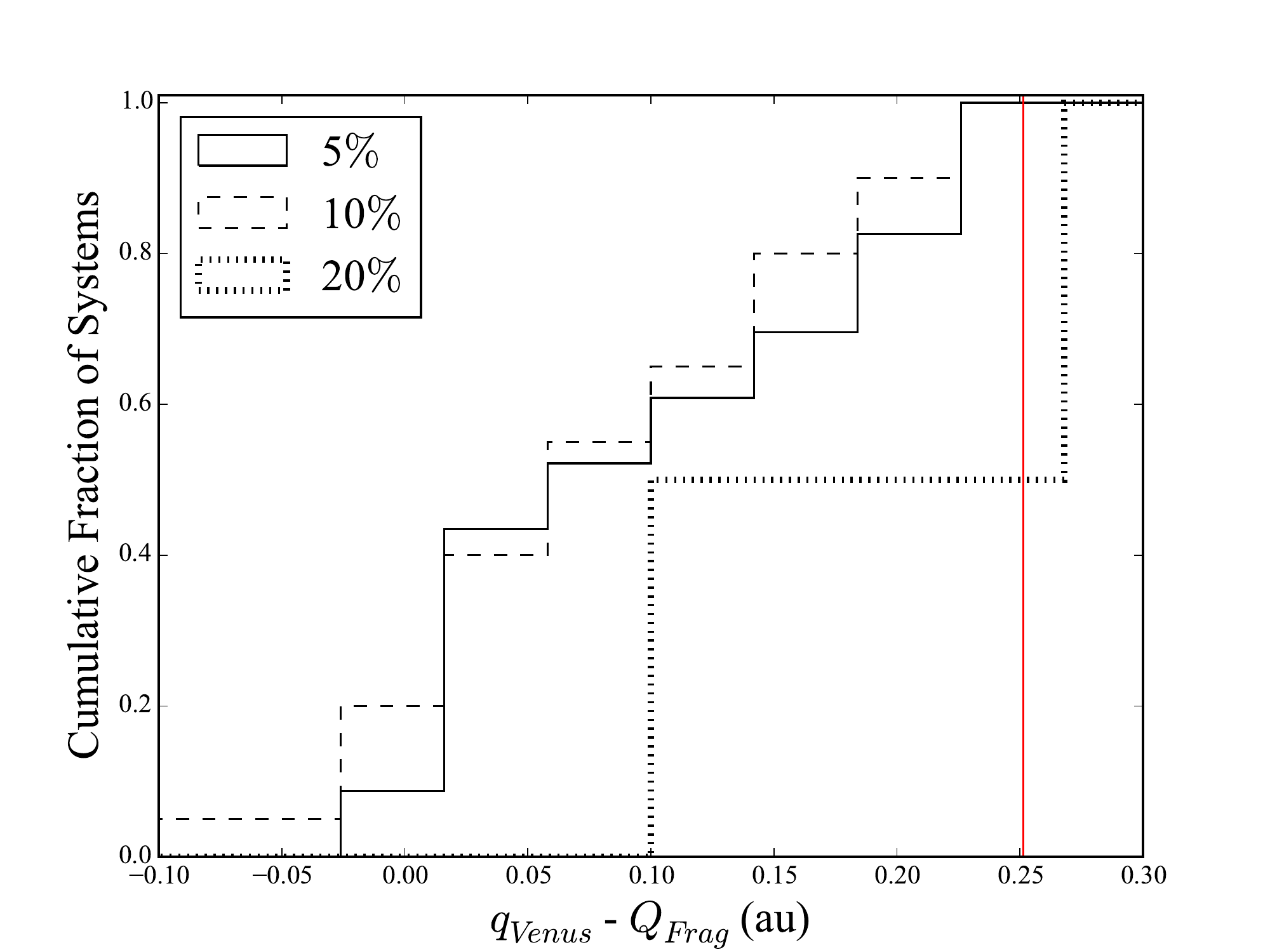}
	\qquad
	\includegraphics[width=.51\textwidth]{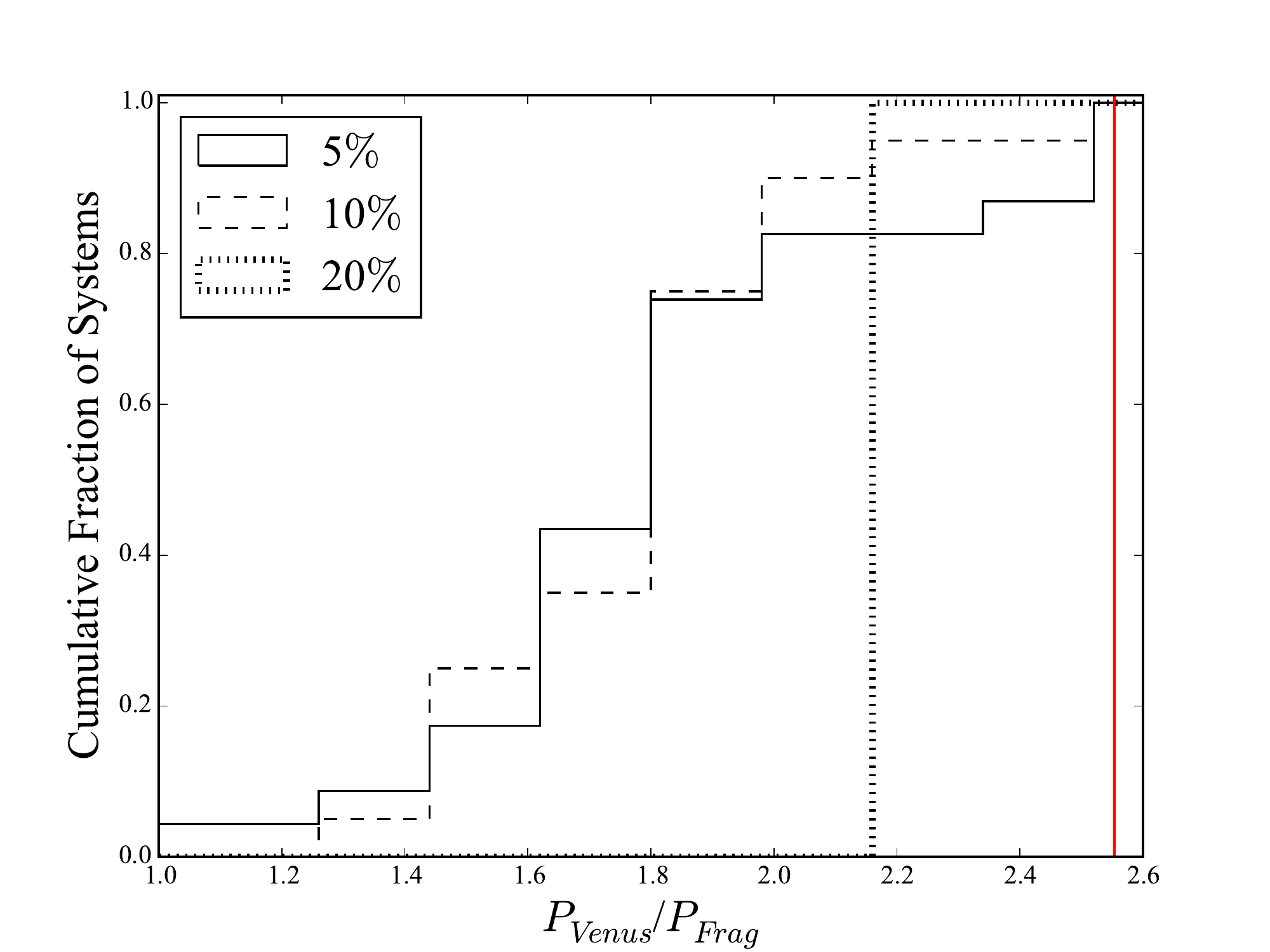}
	\caption{Cumulative distribution of the difference between the perihelion of Venus and aphelion of fragments (top panel), and relative period ratios (bottom panel), in all simulation batches after 1 Gyr of evolution.  The red vertical lines correspond to the solar system values for Mercury and Venus.  The different line styles denote the fragment ejection speed (percentage greater than the mutual escape speed; denoted as subsets a, b and c in table \ref{table:ics}, column 5).}
	\label{fig:qdelt}
\end{figure}

A common outcome of fragmenting collisions (particularly those of the hit-and-run variety) is the re-accretion of ejected fragments \citep{chambers13}.  This is particularly the case when the velocity vectors of the projectile and target objects are near parallel.  In this scenario the fragments are ejected along a plane nearly parallel to the orbit of the target particle; and are still on orbits where they heavily interact with this remnant.  Because the fragments have lower masses than the original projectile object, the subsequent collisions are much more likely to be totally accretionary \citep{lands12}.  We attempt to replicate Mercury and Venus' modern orbital offset by boosting the velocity of escaping fragments.  We find that this only slightly limits the probability of fragment re-accretion.  In such a scenario, the initial two velocity vectors are far from parallel, and the fragments are often ejected on to highly excited orbits.  Over the subsequent billion years of evolution, these bodies can be further excited by chaotic interactions with the other planets \citep{laskar97,laskar09,clement17} to the point where they are either ejected from the system or collide with the Sun.  In general, increasing the fragment ejection velocity leads to a far lower chance of immediate re-accretion by the target body.  However, this difference becomes statistically insignificant when the integration is extended to 1 Gyr because of the longer timescales of re-accretion for high velocity fragments.  Additionally, increasing the ejection speed leads to about a factor of two increase in the chance of loss due to collision with the Sun or ejection from the system.  Figure \ref{fig:qdelt}  plots the Mercury-Venus dynamical offset for all fragments that finish the 1 Gyr integration sunward of Venus in the same manner as figure \ref{fig:qdelt_form}.  Interestingly, though the subsets that fix the fragment ejection velocity at 20$\%$ greater than the mutual escape velocity provide the best matches to the actual solar system, the total sample of such objects (2) is a full order of magnitude smaller than that of the other subsets.  Since all three subsets include roughly an equal number of simulations, we suspect this is due to the fact that subset c (table \ref{table:ics}, column 5) simulations are twice as likely to lose fragments via merger with the Sun than are subsets a or b.  Thus ejecting a fragment with sufficient energy to subsequently scatter off and dynamically separate from the initial target particle also implies a greater chance of losing the fragment by other means.

\begin{table*}
\centering
\begin{tabular}{c c c c c c c}
\hline
Scenario & $a_{o}$ (au) & $e_{o}$ & $i_{o}$ ($^{\circ}$) & $v_{frag}/v_{esc}$ & $N_{sim}$ & $N_{frag}$ \\
\hline
1 & 0.43-0.45 & 0.25-0.5 & 0-1 & (a)1.05,(b)1.1,(c)1.2 & 5472 & 235 \\	
2 & 0.85-0.87 & 0.25-0.5 & 0-1 & (a)1.05,(b)1.1,(c)1.2 & 3286 & 1956 \\	
3 & 0.34-0.36 & 0.25-0.5 & 0-1 & (a)1.05,(b)1.1,(c)1.2 & 9840 & 62\\	
4 & 0.43-0.45 & 0.6-0.75 & 30-40 & (a)1.05 & 569 & 113 \\
\hline	
\end{tabular}
\caption{Summary of initial conditions for simulations to produce Mercury as suggested by \citet{asphaug14}.  The columns are (1) the scenario number, (2-4) the semi-major axis, eccentricity and inclination ranges from which the projectiles orbit is selected, (5) the fragment ejection velocity with respect to the mutual two-body escape velocity (6) the total number of integrations performed, and (7) the total number of integrations where fragmenting collisions occurred.}
\label{table:ics}
\end{table*}

\subsubsection{Terrestrial AMD}

\begin{figure}
	\centering
	\includegraphics[width=.51\textwidth]{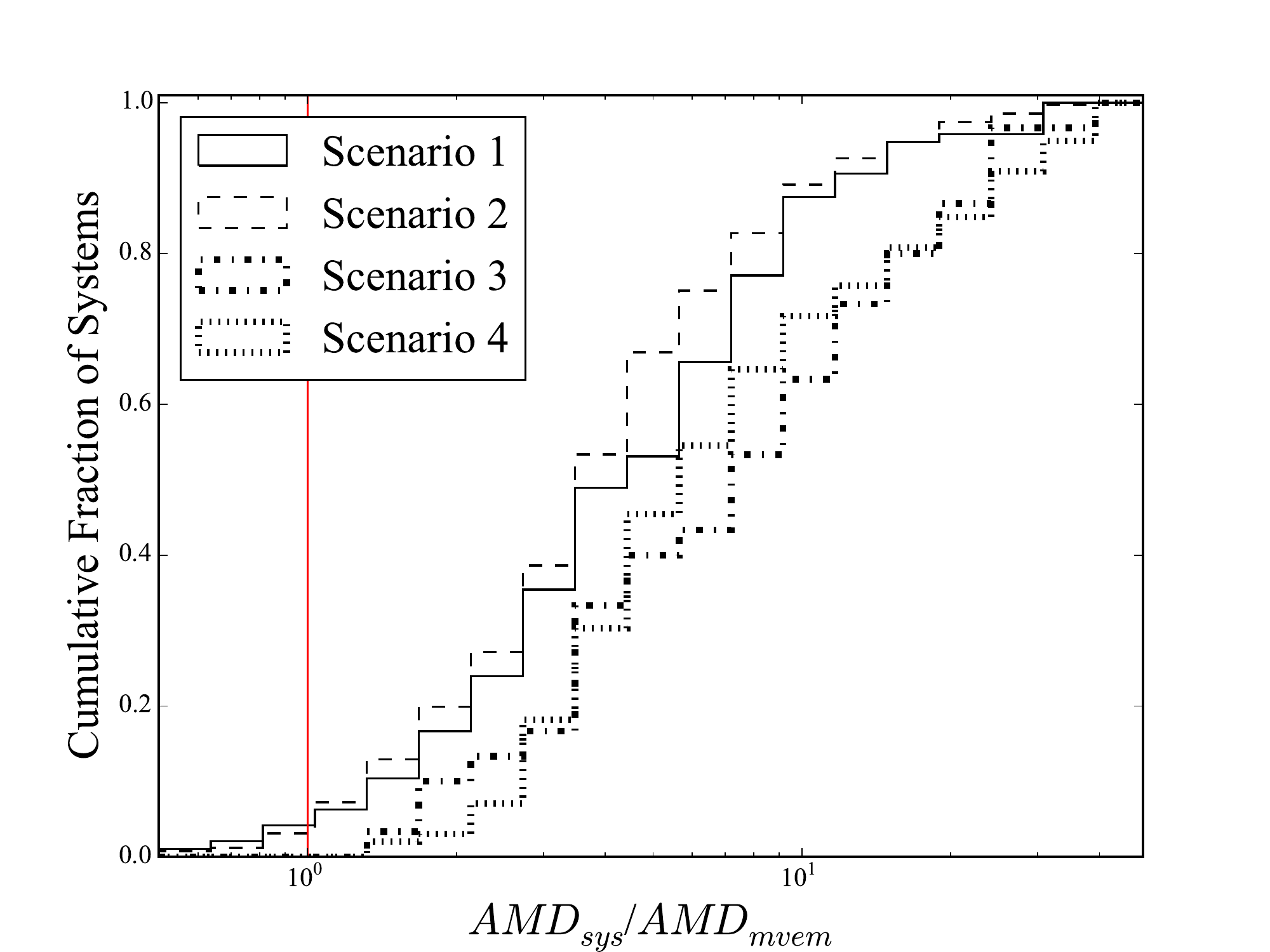}
	\caption{Cumulative distribution of system AMDs after 1 Gyr of evolution, normalized to the solar system value for Mercury, Venus, Earth and Mars.  The different line styles represent the different initial projectile orbital parameter space tested (table \ref{table:ics}).  The red vertical line corresponds to the solar system value.}
	\label{fig:amd}
\end{figure}

Nearly all modern simulations of planet formation in the inner solar system struggle to replicate the extremely low eccentricities and inclinations of the solar system's four terrestrial planets.  In fact, when integrated over Gyr timescales, the orbits of all the  planets except Mercury usually stay remarkably low \citep{quinn91,laskar09}.  For the formation of Mercury by energetic collision model to be viable, the resulting terrestrial system must be appropriately dynamically cold.  In figure \ref{fig:amd}, we plot the cumulative distribution of AMDs across all four of our scenarios.  This final distribution of AMDs should be taken in context with the fact that our simulation set-ups are quite idealized.  If the Mercury-forming impact occurred during the later stages of planet formation, a population of small bodies would have still existed to damp out the excited orbits of the young planets via dynamical friction.  Nevertheless, our results indicate that it is unlikely for a system to undergo such a violent dynamical process and finish in a state similar to the solar system.  The fact that the solar system's AMD is outside the range of the values for scenarios 3 and 4 (table \ref{table:ics}) does speak against these initial projectile orbits.  In particular, no scenario 4 systems finish the 1 Gyr evolution with an AMD within a factor of two of the solar system.  

\citet{jackson18} prefer the multiple hit-and-run setup of \citet{asphaug14} from a probabilistic perspective, however the majority of their favored pre-impact orbits have extremely high inclinations ($\sim$20-60$^{\circ}$).  In the actual solar system, the orbits of all the terrestrial planets except Mercury (e=0.20, i=6.3$^{\circ}$) are nearly circular and co-planar (all three have inclinations less than 2.5$^{\circ}$, Venus and Earth have eccentricities less than 0.02).  For this reason, many studies of early solar system giant impacts only consider co-planar projectiles \citep{quarles15,kaibcowan15}.  In order for a high-inclination impactor scenario to be viable, some mechanism is needed to dissipate the non-perpendicular components of angular momentum delivered by a several-Mars-massed projectile in order to keep the final orbits in the inner solar system dynamically cold.  As a point of reference, only 5 of the terrestrial objects (a $<$ 2.0 au, m $>$ 0.1 $M_{\oplus}$) in our 360  complete planet formation simulations attain inclinations larger than $20^{\circ}$ (the highest being $34^{\circ}$).  This indicates that it is unlikely that the hypothetical Mercury forming impactor could have originated on a highly excited, non-coplanar orbit.

\begin{figure}
	\centering
	\includegraphics[width=.51\textwidth]{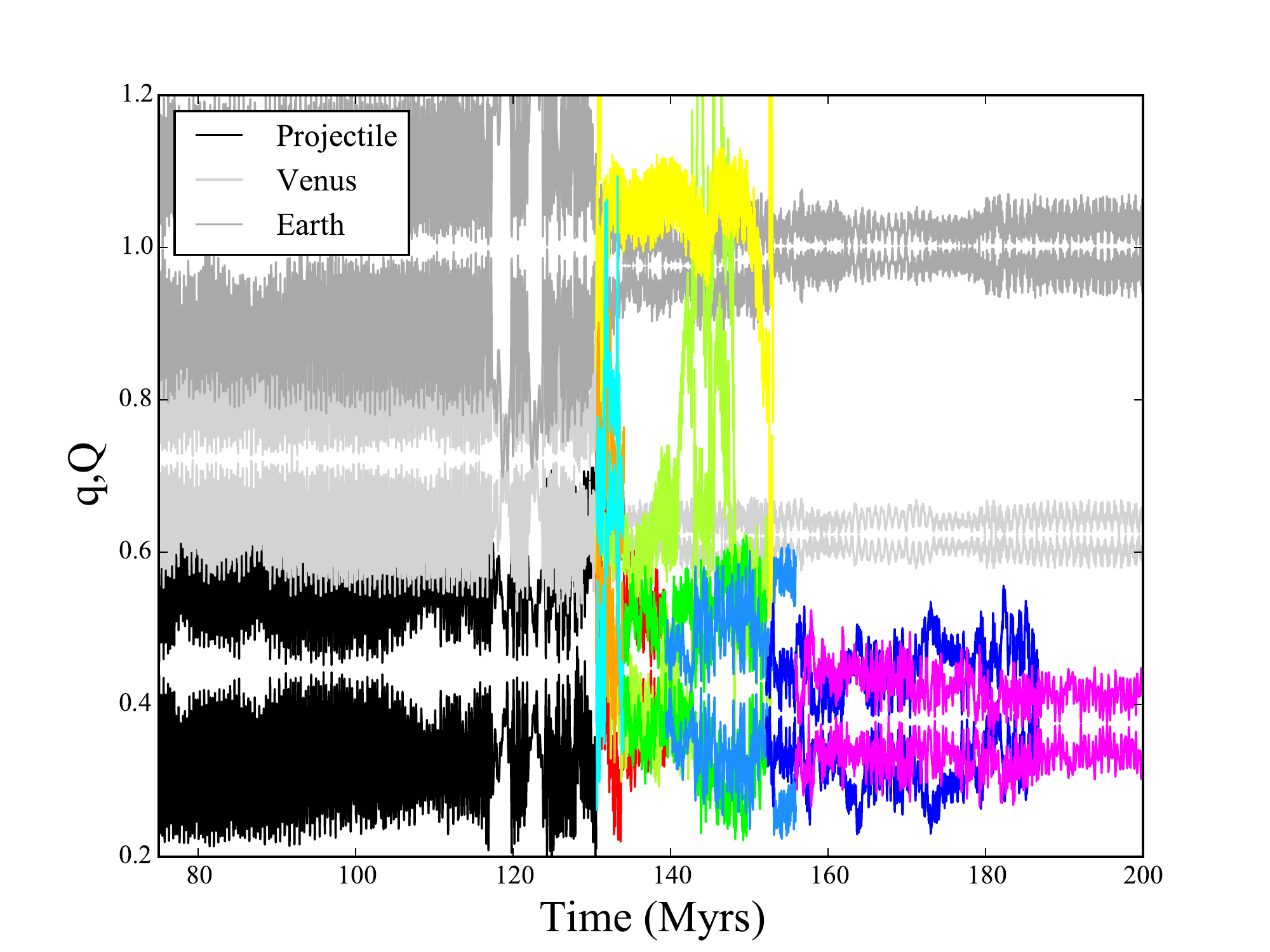}
	\caption{Example evolution of a successful simulation in the scenario 1b set (table \ref{table:ics}, column 5) .  The perihelia and aphelia are plotted for each body in the simulation.  After the projectile (color coded black) smashes into the proto-Venus at 131 Myr, five fragments are ejected.  Four more fragments are formed over the next 25 Myr in four different collisions between the original fragments.  Each fragment is coded a different non-greyscale color.}
	\label{fig:qaq}
\end{figure}

Less than 1$\%$ of simulations in our best subset (scenario 1; table \ref{table:ics}), finish with four terrestrial planets, a system AMD less than twice $AMD_{MVEM}$, and $q_{Venus}$ - $Q_{Frag}$ within $\sim$20$\%$ of the solar system value.  An example of a successful system's evolution is plotted in figure \ref{fig:qaq}.  Initially, the projectile is interacting heavily with the proto-Venus, quickly exciting both objects.  Through repeated close encounters, Venus' eccentricity is excited to $\sim$0.10, and it continues to oscillate between $\sim$0.01 and 0.20 over the next 125 Myr.  Since Venus' excited orbit continually brings it in close proximity to Earth, its eccentricity excitation quickly bleeds to Earth via stochastic diffusion.  Eventually, the projectile smashes into the proto-Venus at a velocity of 1.93 times the mutual escape velocity.  This collision ejects five fragments, each with a mass of $\sim$0.032 $M_{\oplus}$.  4 Myr later, Venus absorbs one of the initial fragments.  Two of the fragments are quickly scattered on to orbits where they heavily interact with Earth, and the other two undergo a series of hit-and-run collisions with one another and Venus.  Through this process, four additional fragments are produced.  Over the next 10 million years, Earth excites the eccentricity of the outer two fragments, placing them on orbits where they eventually merge with Venus.  At 157 Myr, there are just 2 remaining fragments in the system, both interior to Venus' orbit.  These two bodies undergo a series of three hit-and-run collisions before they finally merge at 187 Myr.  During this sequence of repeated hit-and-run collisions with other fragments, the final ``Mercury'' analog dissipates angular momentum and finishes on an orbit that is sufficiently dynamically separated from Venus.

Thus we cannot rule out the \citet{asphaug14} giant impact scenario as dynamically incompatible with Mercury's current orbit.  However, our simplified simulations indicate that it represents a rare and unlikely pathway for Mercury's formation.

\section{Conclusions}

In this paper we presented a dynamical analysis of the various avenues proposed for Mercury's formation utilizing an N-body code that includes the effects of collisional fragmentation.  A major limitation of our work is that the integrator must cap the total number of fragments by setting a MFM in order to maintain a reasonable number of bodies in the calculation.  

In the first part of our study, we utilized the largest sample of complete simulations of terrestrial planet formation that include the effects of collisional fragmentation to search for objects with high CMFs boosted by repeated hit-and-run collisions.  While the very large majority of the planets we produce differ from Mercury in terms of their orbits, masses and CMFs, this sample also provides numerous objects with compositions similar to Mercury.  90$\%$ of all our complete simulations of terrestrial planet formation finish with a high CMF object (CMF$>$0.5).  We find that our annulus and instability sets produce the most high CMF objects.  Our instability simulations are most efficient at generating the largest high CMF objects, and the annulus runs yielded better matches to the terrestrial system as a whole.  Depending on the particular initial conditions of a simulation, we find that planets with similar masses and orbits to Mercury from 1-15$\%$ of the time.  However, only one of our 360 simulations generated a Mercury analog with the proper mass, orbit and CMF within a larger terrestrial architecture that matches the real one.

Additionally, we performed a large suite of simulations designed to replicate the collisional scenario of \citet{asphaug14}.  Our results indicate that such a violent collision occurring late in the giant impact phase represents a very low-likelihood scenario for Mercury's origin.  In particular, replicating the present dynamical separation between Mercury and Venus proves challenging.  Increasing the velocity of escaping fragments to $\sim$20$\%$ greater than the mutual escape velocity can help in attaining this separation, but only to a minor degree.  We also conclude that highly excited, non-co-planar initial projectile orbits similar to those proposed in \citet{jackson18} are unlikely.  In particular, the final systems of planets in this scenario  systematically fail to match the solar system's low AMD.  Despite all the efforts made, forming Mercury continues to be a major challenge for terrestrial planet formation models.

\section*{Acknowledgments}

We thank Sean Raymond and Rogerio Deienno for insightful comments and suggestions that greatly enhanced the quality of this manuscript.  This material is based upon research supported by the Chateaubriand Fellowship of the Office for Science and Technology of the Embassy of France in the United States.  M.S.C. and N.A.K. thank the National Science Foundation for support under award AST-1615975.  This research is part of the Blue Waters sustained-petascale computing project, which is supported by the National Science Foundation (awards OCI-0725070 and ACI-1238993) and the state of Illinois. Blue Waters is a joint effort of the University of Illinois at Urbana-Champaign and its National Center for Supercomputing Applications \citep{bw1,bw2}.  Further computing for this project was performed at the OU Supercomputing Center for Education and Research (OSCER) at the University of Oklahoma (OU).  Additional analysis and simulations were done using resources provided by the Open Science Grid \citep{osg1,osg2}, which is supported by the National Science Foundation award 1148698, and the U.S. Department of Energy's Office of Science.  Several sets of simulations were managed on the Nielsen Hall Network using the HTCondor software package: https://research.cs.wisc.edu/htcondor/.

\bibliographystyle{apj}
\newcommand{\sci}{$Science$ }
\bibliography{mercury.bib}
\end{document}